\documentclass[prl,twocolumn,preprintnumbers,amsmath,amssymb,superscriptaddress]{revtex4-1}

\usepackage{graphicx}
\usepackage{amsmath,amsfonts,amsthm,amssymb}
\usepackage{physics}
\usepackage{float}
\usepackage{dcolumn}
\usepackage{xcolor}

\begin{document}

\title{The effect of curvature on the diffusion of colloidal bananas} 
\author{Justin-Aurel Ulbrich}
\affiliation{Department of Chemistry, Physical and Theoretical Laboratory, University of Oxford, South Parks Road, Oxford OX1 3QZ, United Kingdom}
\affiliation{Department of Materials, ETH Z\"{u}rich, 8093 Zurich, Switzerland}
\author{Carla Fern\'{a}ndez-Rico}
\email{carla.fernandezrico@mat.ethz.ch}
\affiliation{Department of Chemistry, Physical and Theoretical Laboratory, University of Oxford, South Parks Road, Oxford OX1 3QZ, United Kingdom}
\affiliation{Department of Materials, ETH Z\"{u}rich, 8093 Zurich, Switzerland}
\author{Brian Rost}
\affiliation{Department of Physics and Institute for Soft Matter Synthesis and Metrology, \\Georgetown University, Washington, DC 20057.}
\author{Jacopo Vialetto}
\affiliation{Department of Materials, ETH Z\"{u}rich, 8093 Zurich, Switzerland}
\author{Lucio Isa}
\affiliation{Department of Materials, ETH Z\"{u}rich, 8093 Zurich, Switzerland}
\author{Jeffrey S. Urbach} 
\email{urbachj@georgetown.edu}
\affiliation{Department of Physics and Institute for Soft Matter Synthesis and Metrology, \\Georgetown University, Washington, DC 20057.}
\author{Roel P. A. Dullens}
\email{roel.dullens@ru.nl}
\affiliation{Department of Chemistry, Physical and Theoretical Laboratory, University of Oxford, South Parks Road, Oxford OX1 3QZ, United Kingdom}
\affiliation{ Institute for Molecules and Materials, Radboud University,\\ Heyendaalseweg 135, 6525 AJ Nijmegen, The Netherlands}

\begin{abstract}
Anisotropic colloidal particles exhibit complex dynamics which play a crucial role in their functionality, transport and phase behaviour. In this work, we investigate the two-dimensional diffusion of smoothly curved colloidal rods -- also known as colloidal bananas -- as a function of their opening angle, $\alpha$. 
We measure the translational and rotational diffusion coefficients of the particles with opening angles ranging from $0^{\circ}$ (straight rods) to nearly $360^{\circ}$(closed rings). 
In particular, we find that the anisotropic diffusion of the particles varies non-monotonically with their opening angle and that the axis of fastest diffusion switches from the long to the short axis of the particles when $\alpha>180^{\circ}$.
 We also find that the rotational diffusion coefficient of nearly closed rings is approximately an order of magnitude higher than that of straight rods of the same length. Finally, we show that the experimental results are consistent with Slender Body Theory, indicating that the dynamical behavior of the particles arises primarily from their local drag anisotropy. These results highlight the impact of curvature on the Brownian Motion of elongated colloidal particles, which must be taken into account when seeking to understand the behaviour of curved colloidal particles.
\end{abstract}

\maketitle

The Brownian motion of particles suspended in a fluid is a classic topic of study, describing the erratic, random motion of micron-sized particles colliding with the surrounding solvent molecules \cite{Einstein1905,Poon2017,BrMoHangii2005}.
This type of motion has huge implications in the self-assembly, phase behaviour and transport properties of colloidal systems \cite{BookColloid1s,Boles2016rev,Mano2015,Poon2004}, and can be also used as a powerful tool to locally measure the rheology of complex fluids, such as the cell interior \cite{Wirtz2009,Hameed2012}.
Firstly discovered by Brown \cite{Brown1828}, and later formalized by Einstein \cite{Einstein1905}, the extent of Brownian motion is described by the diffusion coefficient, $D$, which for a sphere depends solely on the sphere’s radius, $R$, the temperature, $T$, and viscoelasticty of the medium \cite{MasonWeitz1995}. In the case of a Newtonian fluid, the diffusion coefficient of a sphere is related to the hydrodynamic drag $\xi$ of the particle by famous the Stokes-Einstein relation, $D = k_{B}T \xi^{-1}$, which can be experimentally measured from the mean square displacement (MSD) of the particles in a solvent \cite{Einstein1905,BookColloid1s,Poon2004}.

Introducing shape anisotropy in the particles has a dramatic impact on the way they diffuse. In fact, measuring the diffusion of non-spherical particles is an area of long-standing and ongoing research \cite{Brenner:1967wr, Wegener1981, Tirado:1984vu, Doi:1988ug, Han:2006wb, Han:2009vi, Padding:2010tf, Chakrabarty2013, Chakrabarty2014, Delong2015, Yang:2017ux, Roosen-Runge:2021wo,VerweijPRR2020,Verweij2021}, as many relevant microscopic building blocks, such as functional nanoparticles, protein fibrils and colloidal molecules, are far from isotropic. The challenging step in this process is to  determine the diffusion coefficient of the particles along their different axes and measure the coupling between their translational and rotational motion \cite{Kraft2013,Brenner:1967wr}. To address this, a laboratory and body frame coordinate systems are typically used to separately measure the MSDs along the directions corresponding to particles' axes and decouple their rotational and translational motion \cite{Han:2006wb,Zheng:2010wo,Chakrabarty2014}. For high-aspect-ratio uniaxial particles, such as colloidal ellipsoids and rods, the diffusion tensor is symmetric, meaning that there is no coupling between rotational and translational motion, and the diffusion along the long axis of the particles is higher than perpendicular to it \cite{Tirado:1984vu, Doi:1988ug,Han:2006wb, Han:2009vi, Padding:2010tf, Roosen-Runge:2021wo}. For less symmetric shapes, such as biaxial particles, e.g. colloidal boomerangs \cite{Chakrabarty2013,Chakrabarty2014, Delong2015}, a similar trend is observed with higher diffusivity along the long axis. Nonetheless, the diffusion tensor is more complex and comprises off-diagonal elements accounting for translational-rotational coupling of the dynamics \cite{Chakrabarty2013,Chakrabarty2014,Delong2015}. 

\begin{figure*}[t]
\includegraphics[width=1\textwidth]{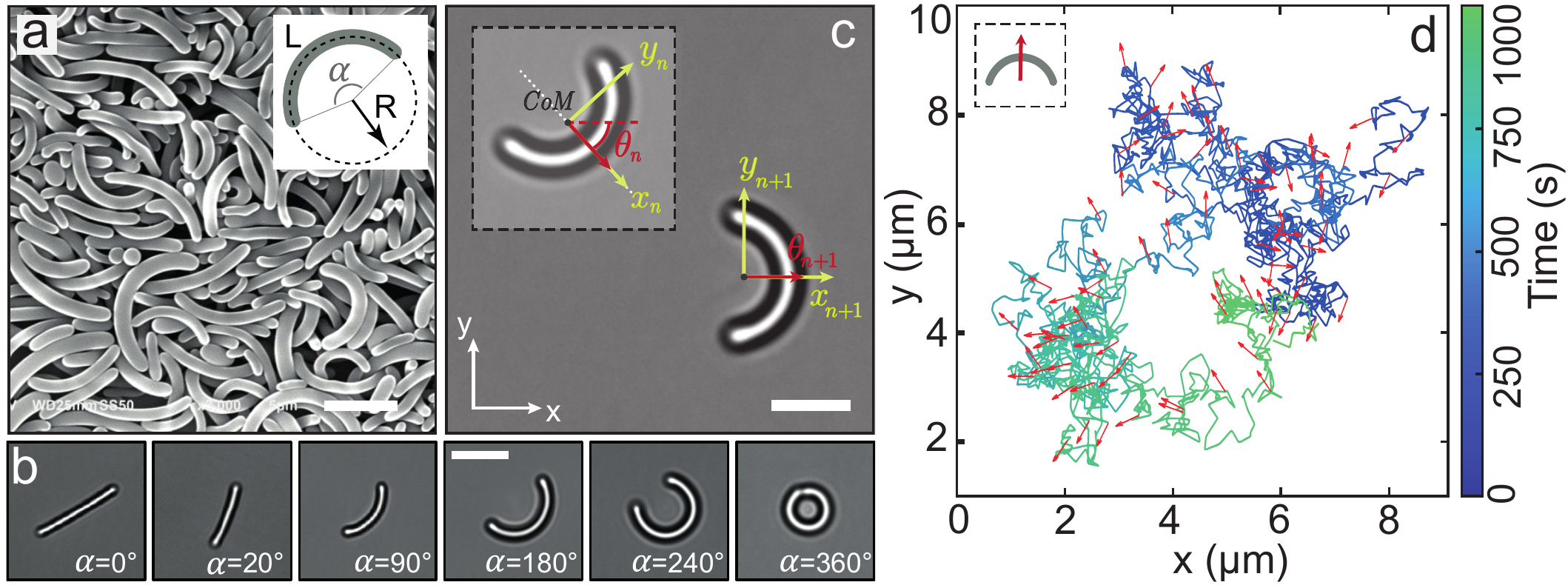}
\caption{Diffusion of colloidal SU-8 banana-shaped particles in two dimensions. (a) Typical scanning electron microscopy image of SU-8 colloidal bananas with an average length, $L$, opening angle, $\alpha$, and radius of curvature, $R$ (see inset). (b) Optical microscopy images of colloidal bananas with increasing opening angle, $\alpha$. (c) Optical microscopy image of a colloidal banana undergoing Brownian motion in the $xy$ plane in two different time points. The lab frame axes ($x$ and $y$, white) are fixed, while the body frame axes (x$_{n}$ and y$_{n}$ yellow) evolve with particle's motion, where $n$ is the frame number. We track the displacement of the Centre of Mass (CoM) along x$_{n+1}$ and y$_{n+1}$, and also their orientation, $\theta_{n}$, defined as the angle between the $x$-axis and the $x_{n}$-axis. The red arrow and white dotted line are the orientation vector of the particles and their symmetry axis, respectively. (d) Typical trajectory and orientation field  of a diffusive colloidal banana over time. All scale bars are 5 $\mu$m. }
\label{setup}
\end{figure*}

Recent advances in colloid synthesis have allowed to not only produce highly anisotropic particles with different symmetries \cite{Hueckel2021,Glotzer:2007uo}, but also particles with a smooth and tunable curvature. This is the case of our recently developed system of \emph{colloidal SU-8 bananas} \cite{Fernandez-Rico:2020wu}, whose curvature is characterized by their opening angle, $\alpha$ (see Fig.~1a). These polymer particles have already given valuable insight into the importance of $\alpha$ in the structure of banana-shaped liquid crystals \cite{Fernandez-Rico:2020wu,Fernandez-Rico:2021uw}, but might also serve as model systems for other curved colloidal systems, such as bacteria \cite{Persat2014,Schuech2019} or curvature-mediating proteins \cite{Qualmann2011,Shen2016}.
This highlights the importance of uncovering their dynamical behaviour as a function of their opening angle, as previous studies have either focused on the dynamics of similar particle shapes but with fixed opening angles \cite{Chakrabarty2013} or on flexible colloidal chains with dynamic opening angles \cite{VerweijPRR2020,Verweij2021}.\\

In this work, we investigate the two-dimensional diffusion of SU-8 colloidal bananas sedimented onto a solid surface, as a function of their opening angle $\alpha$.
To this end, we use particles with $\alpha$ ranging from 0$^{\circ}$ (straight rods) to nearly 360$^{\circ}$ (closed rings), and track their centres of mass in the $xy$ plane over time using optical video microscopy (see Figure 1). We find that the translational diffusion coefficient along  the long axis of the particles, $D_{yy}$, is $\sim$2.4 times higher than that of the short axis, $D_{xx}$, for small $\alpha$, consistent with previous work on elipsoids \cite{Han:2009vi}. When $\alpha$ increases, we observe that the ratio between the coefficients approaches unity, as also expected for closed ring-like objects. However, for $\alpha > 180^{\circ}$, we find a regime where the axis of fastest diffusion switches from the long to the short axis of the particles. In addition, we observe that the rotational diffusion coefficient, $D_\theta$, increases dramatically with increasing $\alpha$, and is about an order of magnitude larger for a nearly-closed ring than for a rod-like particle of the same length. Finally, we find that our measurements are surprisingly well described by analytic results of the simplest form of Slender Body Theory (SBT) \cite{Gray1955,Zhang.2013fzd,Rost2020}, which allows us to conclude that the dynamical behaviour of the particles primarily arises for the local drag anisotropy along the particles, with longer range hydrodynamic couplings playing a secondary role.

A typical scanning electron microscopy image of the SU-8 colloidal bananas used for our diffusion experiments is shown in Figure \ref{setup}a (see synthesis details in \cite{Fernandez-Rico:2020wu}). The shape of the particles is described by a circular arc with a length $L$, opening angle $\alpha$, and radius of curvature $R$ (see inset of Figure \ref{setup}a). 
Our colloidal bananas, which have an approximate diameter of 1$\mu$m, are polydisperse in size with lengths ranging from 8 to 12~$\mu$m ($L/D\sim10$) and opening angles ranging from 0$^{\circ}$ to 360$^{\circ}$ (see Figure 1b). 
We typically prepare very dilute particle suspensions in water (particle volume fraction $\phi<$0.01), and let the particles sediment towards the bottom surface of our sample cell. In order to confine the motion of the particles in two dimensions, we add a small amount of depletant (1~mg/mL of aqueous xanthan solution) which introduces a weak depletion attraction ($\sim$10$k_{B}T$) between the particles and the cell wall \cite{Lekkerkerker1992}, and suppresses any out-of-plane motion of the particles. The resulting quasi two-dimensional system is shown in Figure 1c, which shows an optical microscopy image of our system of colloidal bananas moving in the $xy$~plane. 

\begin{figure*}[t]
\includegraphics[width=\textwidth]{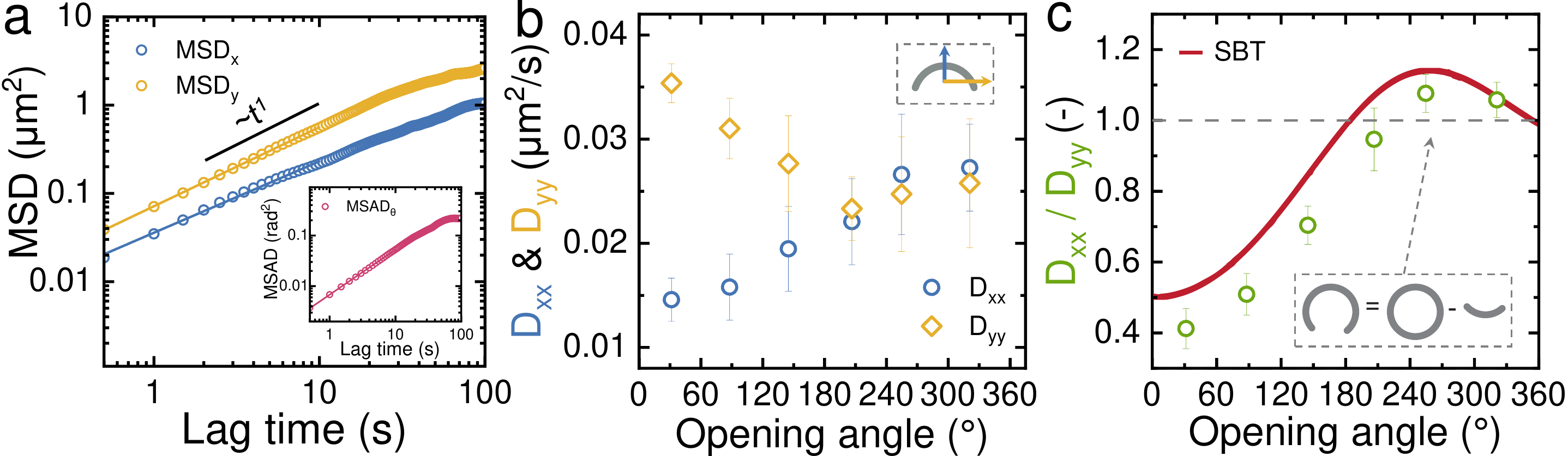}
\caption{Translational diffusion coefficients as a function of the opening angle, $\alpha$, of the particles. (a) Typical mean squared displacement of a colloidal banana with $\alpha=100^{\circ}$ (note that the error bars are smaller than the symbols). The inset shows the angular MSD (MSAD). All curves show diffusive behaviour ($\propto$ t) for lag times below 10~s. The lines show the best linear fit for short times where the fitting error is $<$2\%. (b) Translation diffusion coefficients as a function of $\alpha$, along the short axis, $D_{xx}$, and the long axis, $D_{yy}$, of the particles. Each bin of 60$^{\circ}$ is averaged over $\sim$15 particles with a mean length of 10$\pm 2 \mu$m. The error bars represent the standard deviation of the mean value. (c) Ratio of  $D_{xx}/D_{yy}$ as a function of $\alpha$. The solid line shows the analytical result from Slender Body Theory (SBT) and the inset a sketch of a colloidal banana with a large opening angle. } 
\label{translational}
\end{figure*}

Next, we track the position and orientation of the particles over time, by measuring their centre of mass (CoM) and orientation displacements using video optical microscopy (see resulting trajectory in Figure 1d). We extract the particles' translational and rotational diffusion coefficients by using particle tracking algorithms based on Chakrabaty's work \cite{Chakrabarty2013,Chakrabarty2014}. The key parameters used for such measurements are shown in Figure 1c. In short, we use two coordinate systems, the laboratory frame, which is fixed over time (see white axes in Fig.~1c), and a continuous body frame, which is centred on the centre of mass of the particle and moves with the particle's motion (see yellow axes in Fig.~1c). Note that the body frame is always chosen such that its x-axis, $x_{n}$, is parallel to the particle orientation vector (see red arrow in Fig.~1c), which lies on the symmetry axis of colloidal bananas (see dotted line Fig.~1c).

The centre of mass displacements in the body frame, $\boldsymbol{\delta r}(t)^{BF}$, are obtained using a rotational coordinate transformation of the displacements in the lab frame, $\boldsymbol{\delta r}(t)^{LF}$ following $\boldsymbol{\delta r}(t)^{BF}=\boldsymbol{R}(\theta)\cdot\delta\boldsymbol{r}(t)^{LF}$, where $\boldsymbol{\delta r}(t) = ({\delta x_{n}}, {\delta y_{n}})$ and $\boldsymbol{R}(\theta)$ is the rotational transformation matrix \cite{Chakrabarty2014}. Here, $\theta$ is the particle orientation angle (see angle in Figure 1c), which is defined as the average angle between the final and initial orientations in the displacement of the particle. In this framework, the translation diffusion of the particle in the short axis, D$_{xx}$, is decoupled from the rotational diffusion D$_{\theta}$ due to the symmetry of the particles \cite{Chakrabarty2013,Chakrabarty2014}. On the contrary, the translational diffusion along the long axis, D$_{yy}$, remains coupled to the rotational diffusion regardless of the definition of $\theta$, unless the centre of hydrodynamics is used as the tracking point \cite{Chakrabarty2013,Chakrabarty2014}. Nonetheless, as we discuss later in the text, this coupling is negligible as the distance between the CoM and CoH in our system is very small (see Figure S2a). In the rest of the work, we use the methods explained above to systematically study how the translational and rotational diffusion coefficients are affected by the opening angle of the particles. 

\begin{figure*}
	\centering
	\includegraphics[width=\textwidth]
{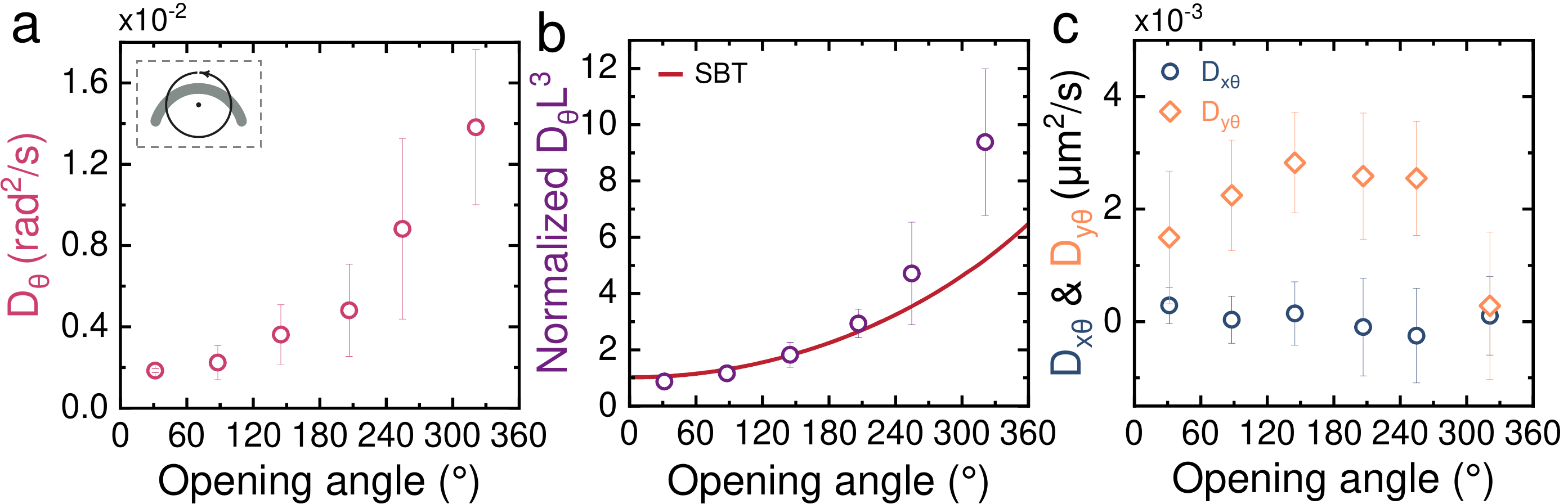}
\caption{(a) Rotational diffusion and coupling coefficients  for colloidal bananas as a function of their opening angle. (a) Rotational diffusion, $D_{\theta}$, as a function of $\alpha$. Each bin of 60$^{\circ}$ is averaged over approximately 15 particles with a mean length of 10$\pm 2 \mu$m. (b) $D_{\theta}$ times $L^3$ for each particle, normalized by the average value of $D_{\theta} L^3$ measured for rods (N=20). Solid line: Analytical result from Slender body Theory. (c) Translational-Rotational coupling diffusion coefficients as a function of the opening angle. While no coupling is observed between rotations and the short-axis of the particles, a non-zero coupling is observed for intermediate opening angles. The error bars represent the standard deviation of the average value.} 
\label{rotational}
\end{figure*}

First, we consider the translational diffusion of the colloidal bananas. Figure 2a shows a representative mean squared displacement (MSD) in the body reference frame of a colloidal banana with $\alpha=100^{\circ}$. All MSD plots show diffusive behaviour for lag times below 10 seconds. From short diffusion times (see SI), we extract the translational diffusion coefficients measured along the short, $D_{xx}$, and long axes of the particles, $D_{yy}$, as a function of the opening angle (see Figure \ref{translational}b). Firstly, we observe that $D_{yy}$ largely exceeds $D_{xx}$ for small opening angles. This behaviour is expected, as for a rod-like particle ($\alpha \sim 0^{\circ}$), the diffusion along the long axis is known to be approximately twice the diffusion along its short axis \cite{Doi:1988ug,Han:2009vi}. For large opening angles, the difference between $D_{xx}$ and $D_{yy}$ decreases, which is also expected as for a ring ($\alpha \sim 360^{\circ}$), $D_{xx}$=$D_{yy}$ due to symmetry reasons. Note that the scatter in the diffusion coefficient values, represented as the standard deviation in the error bars in Figure 2b , originates primarily from the polydispersity ($\sim20\%$) in the particles' length in each bin.
\vspace{0.2cm}

Interestingly, we find that for opening angles above 200$^{\circ}$, the axis of fastest diffusion switches from the long to the short axis of the particles, with $D_{xx}$ exceeding $D_{yy}$ (see Figure 2b, $\alpha > 200^{\circ}$). This effect shows up more clearly in Figure \ref{translational}c, where we plot the ratio of $D_{xx}/D_{yy}$ as a function of the opening angle, and observe that it clearly exceeds unity for 200$^{\circ}< \alpha < 360^{\circ}$. For  $\alpha \sim 360^{\circ}$, $D_{xx}/D_{yy}$ approaches unity again, as expected for a ring. The non-monotonical behaviour of the measured anisotropic diffusion can be explained in terms of drag forces. For a ring, the drag force experienced by the particle is independent of the direction of motion, and therefore $D_{xx}$=$D_{yy}$. However, when a small segment of the ring is removed, the resulting translational diffusion of the broken ring becomes anisotropic with $D_{xx} > D_{yy}$. This is because the section removed has a lower longitudinal drag force than the transverse one ($D_{yy} > D_{xx}$), and therefore the $D_{xx}$ of the broken ring is reduced less than its $D_{yy}$ (see sketch of broken ring in the inset of Fig.~2c).

This behaviour is surprisingly well captured by the simplest form of Slender Body Theory (see solid line in Figure \ref{translational}c), which considers the hydrodynamic behaviour of an infinitesimally thin banana, assuming that the perpendicular drag force of the body is twice the parallel one, and that there are no couplings between the different parts of the body (see SI) \cite{Gray1955,Zhang.2013fzd,Rost2020}. The observed deviations from the experimental results stem from the fact that our particles are obviously not infinitesimally thin, from the presence of depletant and from wall effects. In fact, our experimental results agree with previous experimental studies of confined ellipsoids \cite{Han:2009vi,Zheng:2010wo}, where an enhanced diffusion along the long axis is measured with $D_{xx}/D_{yy}<0.5$ due to the presence of a wall, a behaviour that is not captured by the theoretical values shown here. 
\vspace{0.2cm}

Next, we study the dependence of the rotational diffusion coefficient, $D_{\theta}$, on the opening angle of the particles. As shown in Figure \ref{rotational}a, we find that the rotational diffusivity of the colloidal bananas increases dramatically with increasing opening angle. In fact, if we normalize our results by the average $D_\theta L^3$ measured for 20 straight rods, we find that $D_\theta$ is about an order of magnitude larger for a ring than for a rod of the same length ($D_\theta^{ring}/D_\theta^{rod} \approx 10$, see Figure 3b). Part of this increase can be understood in the context of SBT (see solid line in Figure 3b), which not only considers drag forces but also torques introduced by small rotations of the particles around their centre of mass (see SI). When accounting for these two factors, SBT predicts that $D_\theta^{ring}/D_\theta^{rod} \approx 6.6$, for particles with the same length (see SI). However, this result clearly underestimates our measurements for $D_{\theta}$ at large opening angles (see Figure 3b, $\alpha > 240^{\circ}$), which may be explained by the emergence of hydrodynamic coupling between the different parts of the particles as $\alpha$ increases, or by an enhanced anisotropic diffusion near solid surfaces \cite{Han:2009vi}. These effects may  reduce the rotational drag of the particles and lead to the observed enhanced rotational diffusion for large opening angles.

Finally, we measure the coupling between the translational and rotational diffusion as a function of the opening angle, as shown in Figure 3c. Indeed, we find that such coupling is approximately zero between the short axis and the orientation of the particles, $D_{x \theta}$, for all opening angles \cite{Chakrabarty2014}. We also find that there is a non-zero coupling between the long axis and rotation of the particles, $D_{y \theta}$, which is expected as we are using the centre of mass and not the centre of hydrodynamics as the tracking point \cite{Chakrabarty2014}. Nonetheless, such coupling is nearly zero for small and large opening angles (see Figure 3c for $\alpha \sim 0^{\circ}$ and $\alpha \sim 360^{\circ}$) and an order of magnitude smaller than $D_{yy}/L$ for intermediate opening angles (see Figure S2b), which means that the trends described previously in Figure 2 are indeed not significantly affected by this coupling.

In summary, we have measured the translational and rotational diffusion coefficients of smoothly curved colloidal bananas in two dimensions, as a function of their opening angle, $\alpha$. We find that the typical anisotropic diffusion expected for straight rods decreases with increasing opening angle until $\alpha>180^{\circ}$, at which point the axis of fastest diffusion switches from the long to the short axis of the particle. We also find that the rotational diffusion coefficient increases dramatically as the opening angle increases. We show that both observations are well described by the simplest form of Slender Body Theory, indicating that these effects can be mainly understood by local drag forces along the particle. These results demonstrate key role of curvature on the Brownian motion of elongated particles, and highlights the richness in the diffusive behaviour of colloids with curved shapes, including colloidal molecules, bacteria or functional nanoparticles.\\

Laura Alvarez is thanked for useful discussions. RPAD and CFR acknowledge the European Research Council (ERC Consolidator Grant No. 724834) for financial support. JAU acknowledges financial support from the Fondation Zdenek et Michaela Bakala.
BR and JSU acknowledge support from the National Science Foundation (Grant No. ENG-1907705).\\

JAU and CFR contributed equally to this work.

%


\begin{thebibliography}{39}%
	\makeatletter
	\providecommand \@ifxundefined [1]{%
		\@ifx{#1\undefined}
	}%
	\providecommand \@ifnum [1]{%
		\ifnum #1\expandafter \@firstoftwo
		\else \expandafter \@secondoftwo
		\fi
	}%
	\providecommand \@ifx [1]{%
		\ifx #1\expandafter \@firstoftwo
		\else \expandafter \@secondoftwo
		\fi
	}%
	\providecommand \natexlab [1]{#1}%
	\providecommand \enquote  [1]{``#1''}%
	\providecommand \bibnamefont  [1]{#1}%
	\providecommand \bibfnamefont [1]{#1}%
	\providecommand \citenamefont [1]{#1}%
	\providecommand \href@noop [0]{\@secondoftwo}%
	\providecommand \href [0]{\begingroup \@sanitize@url \@href}%
	\providecommand \@href[1]{\@@startlink{#1}\@@href}%
	\providecommand \@@href[1]{\endgroup#1\@@endlink}%
	\providecommand \@sanitize@url [0]{\catcode `\\12\catcode `\$12\catcode
		`\&12\catcode `\#12\catcode `\^12\catcode `\_12\catcode `\%12\relax}%
	\providecommand \@@startlink[1]{}%
	\providecommand \@@endlink[0]{}%
	\providecommand \url  [0]{\begingroup\@sanitize@url \@url }%
	\providecommand \@url [1]{\endgroup\@href {#1}{\urlprefix }}%
	\providecommand \urlprefix  [0]{URL }%
	\providecommand \Eprint [0]{\href }%
	\providecommand \doibase [0]{http://dx.doi.org/}%
	\providecommand \selectlanguage [0]{\@gobble}%
	\providecommand \bibinfo  [0]{\@secondoftwo}%
	\providecommand \bibfield  [0]{\@secondoftwo}%
	\providecommand \translation [1]{[#1]}%
	\providecommand \BibitemOpen [0]{}%
	\providecommand \bibitemStop [0]{}%
	\providecommand \bibitemNoStop [0]{.\EOS\space}%
	\providecommand \EOS [0]{\spacefactor3000\relax}%
	\providecommand \BibitemShut  [1]{\csname bibitem#1\endcsname}%
	\let\auto@bib@innerbib\@empty
	\bibitem [{\citenamefont {Einstein}(1905)}]{Einstein1905}%
	\BibitemOpen
	\bibfield  {author} {\bibinfo {author} {\bibfnamefont {A.}~\bibnamefont
			{Einstein}},\ }\href@noop {} {\bibfield  {journal} {\bibinfo  {journal}
			{Annalen der Physik}\ }\textbf {\bibinfo {volume} {322}},\ \bibinfo {pages}
		{549} (\bibinfo {year} {1905})}\BibitemShut {NoStop}%
	\bibitem [{\citenamefont {Poon}(2017)}]{Poon2017}%
	\BibitemOpen
	\bibfield  {author} {\bibinfo {author} {\bibfnamefont {W.~C.}\ \bibnamefont
			{Poon}},\ }\href@noop {} {\bibfield  {journal} {\bibinfo  {journal} {The
				Oxford Handbook of Soft Condensed Matter}\ ,\ \bibinfo {pages} {1}} (\bibinfo
		{year} {2017})}\BibitemShut {NoStop}%
	\bibitem [{\citenamefont {Hänggi}\ and\ \citenamefont
		{Marchesoni}(2005)}]{BrMoHangii2005}%
	\BibitemOpen
	\bibfield  {author} {\bibinfo {author} {\bibfnamefont {P.}~\bibnamefont
			{Hänggi}}\ and\ \bibinfo {author} {\bibfnamefont {F.}~\bibnamefont
			{Marchesoni}},\ }\href@noop {} {\bibfield  {journal} {\bibinfo  {journal}
			{Chaos: An Interdisciplinary Journal of Nonlinear Science}\ }\textbf
		{\bibinfo {volume} {15}},\ \bibinfo {pages} {026101} (\bibinfo {year}
		{2005})}\BibitemShut {NoStop}%
	\bibitem [{\citenamefont {Evans}\ and\ \citenamefont
		{Wennerstrom}(1999)}]{BookColloid1s}%
	\BibitemOpen
	\bibfield  {author} {\bibinfo {author} {\bibfnamefont {D.~F.}\ \bibnamefont
			{Evans}}\ and\ \bibinfo {author} {\bibfnamefont {H.}~\bibnamefont
			{Wennerstrom}},\ }\href@noop {} {\enquote {\bibinfo {title} {Colloidal
				domain},}\ } (\bibinfo {year} {1999})\BibitemShut {NoStop}%
	\bibitem [{\citenamefont {Boles}\ \emph {et~al.}(2016)\citenamefont {Boles},
		\citenamefont {Engel},\ and\ \citenamefont {Talapin}}]{Boles2016rev}%
	\BibitemOpen
	\bibfield  {author} {\bibinfo {author} {\bibfnamefont {M.~A.}\ \bibnamefont
			{Boles}}, \bibinfo {author} {\bibfnamefont {M.}~\bibnamefont {Engel}}, \ and\
		\bibinfo {author} {\bibfnamefont {D.~V.}\ \bibnamefont {Talapin}},\
	}\href@noop {} {\bibfield  {journal} {\bibinfo  {journal} {Chemical Reviews}\
		}\textbf {\bibinfo {volume} {116}},\ \bibinfo {pages} {11220} (\bibinfo
		{year} {2016})}\BibitemShut {NoStop}%
	\bibitem [{\citenamefont {Manoharan}(2015)}]{Mano2015}%
	\BibitemOpen
	\bibfield  {author} {\bibinfo {author} {\bibfnamefont {V.~N.}\ \bibnamefont
			{Manoharan}},\ }\href@noop {} {\bibfield  {journal} {\bibinfo  {journal}
			{Science}\ }\textbf {\bibinfo {volume} {349}},\ \bibinfo {pages} {1253751}
		(\bibinfo {year} {2015})}\BibitemShut {NoStop}%
	\bibitem [{\citenamefont {Poon}(2004)}]{Poon2004}%
	\BibitemOpen
	\bibfield  {author} {\bibinfo {author} {\bibfnamefont {W.}~\bibnamefont
			{Poon}},\ }\href {\doibase 10.1126/science.1097964} {\bibfield  {journal}
		{\bibinfo  {journal} {Science}\ }\textbf {\bibinfo {volume} {304}},\ \bibinfo
		{pages} {830} (\bibinfo {year} {2004})}\BibitemShut {NoStop}%
	\bibitem [{\citenamefont {Wirtz}(2009)}]{Wirtz2009}%
	\BibitemOpen
	\bibfield  {author} {\bibinfo {author} {\bibfnamefont {D.}~\bibnamefont
			{Wirtz}},\ }\href@noop {} {\bibfield  {journal} {\bibinfo  {journal} {Annual
				Review of Biophysics}\ }\textbf {\bibinfo {volume} {38}},\ \bibinfo {pages}
		{301} (\bibinfo {year} {2009})}\BibitemShut {NoStop}%
	\bibitem [{\citenamefont {Hameed}\ \emph {et~al.}(2012)\citenamefont {Hameed},
		\citenamefont {Rao},\ and\ \citenamefont {Shivashankar}}]{Hameed2012}%
	\BibitemOpen
	\bibfield  {author} {\bibinfo {author} {\bibfnamefont {F.~M.}\ \bibnamefont
			{Hameed}}, \bibinfo {author} {\bibfnamefont {M.}~\bibnamefont {Rao}}, \ and\
		\bibinfo {author} {\bibfnamefont {G.~V.}\ \bibnamefont {Shivashankar}},\
	}\href@noop {} {\bibfield  {journal} {\bibinfo  {journal} {PLOS ONE}\
		}\textbf {\bibinfo {volume} {7}},\ \bibinfo {pages} {1} (\bibinfo {year}
		{2012})}\BibitemShut {NoStop}%
	\bibitem [{\citenamefont {Brown}(2009)}]{Brown1828}%
	\BibitemOpen
	\bibfield  {author} {\bibinfo {author} {\bibfnamefont {R.}~\bibnamefont
			{Brown}},\ }\href@noop {} {\bibfield  {journal} {\bibinfo  {journal} {The
				Philosophical Magazine}\ }\textbf {\bibinfo {volume} {4}},\ \bibinfo {pages}
		{161} (\bibinfo {year} {2009})}\BibitemShut {NoStop}%
	\bibitem [{\citenamefont {Mason}\ and\ \citenamefont
		{Weitz}(1995)}]{MasonWeitz1995}%
	\BibitemOpen
	\bibfield  {author} {\bibinfo {author} {\bibfnamefont {T.~G.}\ \bibnamefont
			{Mason}}\ and\ \bibinfo {author} {\bibfnamefont {D.~A.}\ \bibnamefont
			{Weitz}},\ }\href@noop {} {\bibfield  {journal} {\bibinfo  {journal}
			{Physical Review Letters}\ }\textbf {\bibinfo {volume} {74}},\ \bibinfo
		{pages} {1250} (\bibinfo {year} {1995})}\BibitemShut {NoStop}%
	\bibitem [{\citenamefont {Brenner}(1967)}]{Brenner:1967wr}%
	\BibitemOpen
	\bibfield  {author} {\bibinfo {author} {\bibfnamefont {H.}~\bibnamefont
			{Brenner}},\ }\href@noop {} {\bibfield  {journal} {\bibinfo  {journal}
			{Journal of Colloid and Interface Science}\ } (\bibinfo {year}
		{1967})}\BibitemShut {NoStop}%
	\bibitem [{\citenamefont {Wegener}(1981)}]{Wegener1981}%
	\BibitemOpen
	\bibfield  {author} {\bibinfo {author} {\bibfnamefont {W.~A.}\ \bibnamefont
			{Wegener}},\ }\href@noop {} {\bibfield  {journal} {\bibinfo  {journal}
			{Biopolymers}\ }\textbf {\bibinfo {volume} {20}},\ \bibinfo {pages} {303}
		(\bibinfo {year} {1981})}\BibitemShut {NoStop}%
	\bibitem [{\citenamefont {Tirado}\ \emph {et~al.}(1984)\citenamefont {Tirado},
		\citenamefont {Mart{\'\i}nez},\ and\ \citenamefont {Torre}}]{Tirado:1984vu}%
	\BibitemOpen
	\bibfield  {author} {\bibinfo {author} {\bibfnamefont {M.~M.}\ \bibnamefont
			{Tirado}}, \bibinfo {author} {\bibfnamefont {C.~L.}\ \bibnamefont
			{Mart{\'\i}nez}}, \ and\ \bibinfo {author} {\bibfnamefont {J.~G. d.~l.}\
			\bibnamefont {Torre}},\ }\href@noop {} {\bibfield  {journal} {\bibinfo
			{journal} {The Journal of Chemical Physics}\ }\textbf {\bibinfo {volume}
			{81}},\ \bibinfo {pages} {2047} (\bibinfo {year} {1984})}\BibitemShut
	{NoStop}%
	\bibitem [{\citenamefont {Doi}\ and\ \citenamefont
		{Edwards}(1988)}]{Doi:1988ug}%
	\BibitemOpen
	\bibfield  {author} {\bibinfo {author} {\bibfnamefont {M.}~\bibnamefont
			{Doi}}\ and\ \bibinfo {author} {\bibfnamefont {S.~F.}\ \bibnamefont
			{Edwards}},\ }\href@noop {} {\emph {\bibinfo {title} {The Theory of Polymer
				Dynamics}}},\ Clarendon Press\ (\bibinfo  {publisher} {Oxford University
		Press},\ \bibinfo {year} {1988})\BibitemShut {NoStop}%
	\bibitem [{\citenamefont {Han}\ \emph {et~al.}(2006)\citenamefont {Han},
		\citenamefont {Alsayed}, \citenamefont {Nobili}, \citenamefont {Zhang},
		\citenamefont {Lubensky},\ and\ \citenamefont {Yodh}}]{Han:2006wb}%
	\BibitemOpen
	\bibfield  {author} {\bibinfo {author} {\bibfnamefont {Y.}~\bibnamefont
			{Han}}, \bibinfo {author} {\bibfnamefont {A.~M.}\ \bibnamefont {Alsayed}},
		\bibinfo {author} {\bibfnamefont {M.}~\bibnamefont {Nobili}}, \bibinfo
		{author} {\bibfnamefont {J.}~\bibnamefont {Zhang}}, \bibinfo {author}
		{\bibfnamefont {T.~C.}\ \bibnamefont {Lubensky}}, \ and\ \bibinfo {author}
		{\bibfnamefont {A.~G.}\ \bibnamefont {Yodh}},\ }\href@noop {} {\bibfield
		{journal} {\bibinfo  {journal} {Science}\ }\textbf {\bibinfo {volume}
			{314}},\ \bibinfo {pages} {626} (\bibinfo {year} {2006})}\BibitemShut
	{NoStop}%
	\bibitem [{\citenamefont {Han}\ \emph {et~al.}(2009)\citenamefont {Han},
		\citenamefont {Alsayed}, \citenamefont {Nobili},\ and\ \citenamefont
		{Yodh}}]{Han:2009vi}%
	\BibitemOpen
	\bibfield  {author} {\bibinfo {author} {\bibfnamefont {Y.}~\bibnamefont
			{Han}}, \bibinfo {author} {\bibfnamefont {A.}~\bibnamefont {Alsayed}},
		\bibinfo {author} {\bibfnamefont {M.}~\bibnamefont {Nobili}}, \ and\ \bibinfo
		{author} {\bibfnamefont {A.~G.}\ \bibnamefont {Yodh}},\ }\href@noop {}
	{\bibfield  {journal} {\bibinfo  {journal} {Physical Review E}\ }\textbf
		{\bibinfo {volume} {80}},\ \bibinfo {pages} {011403} (\bibinfo {year}
		{2009})}\BibitemShut {NoStop}%
	\bibitem [{\citenamefont {Padding}\ and\ \citenamefont
		{Briels}(2010)}]{Padding:2010tf}%
	\BibitemOpen
	\bibfield  {author} {\bibinfo {author} {\bibfnamefont {J.~T.}\ \bibnamefont
			{Padding}}\ and\ \bibinfo {author} {\bibfnamefont {W.~J.}\ \bibnamefont
			{Briels}},\ }\href {\doibase 10.1063/1.3308649} {\bibfield  {journal}
		{\bibinfo  {journal} {The Journal of Chemical Physics}\ }\textbf {\bibinfo
			{volume} {132}},\ \bibinfo {pages} {054511} (\bibinfo {year}
		{2010})}\BibitemShut {NoStop}%
	\bibitem [{\citenamefont {Chakrabarty}\ \emph {et~al.}(2013)\citenamefont
		{Chakrabarty}, \citenamefont {Konya}, \citenamefont {Wang}, \citenamefont
		{Selinger}, \citenamefont {Sun},\ and\ \citenamefont
		{Wei}}]{Chakrabarty2013}%
	\BibitemOpen
	\bibfield  {author} {\bibinfo {author} {\bibfnamefont {A.}~\bibnamefont
			{Chakrabarty}}, \bibinfo {author} {\bibfnamefont {A.}~\bibnamefont {Konya}},
		\bibinfo {author} {\bibfnamefont {F.}~\bibnamefont {Wang}}, \bibinfo {author}
		{\bibfnamefont {J.~V.}\ \bibnamefont {Selinger}}, \bibinfo {author}
		{\bibfnamefont {K.}~\bibnamefont {Sun}}, \ and\ \bibinfo {author}
		{\bibfnamefont {Q.~H.}\ \bibnamefont {Wei}},\ }\href@noop {} {\bibfield
		{journal} {\bibinfo  {journal} {Physical Review Letters}\ }\textbf {\bibinfo
			{volume} {111}},\ \bibinfo {pages} {160603} (\bibinfo {year}
		{2013})}\BibitemShut {NoStop}%
	\bibitem [{\citenamefont {Chakrabarty}\ \emph {et~al.}(2014)\citenamefont
		{Chakrabarty}, \citenamefont {Konya}, \citenamefont {Wang}, \citenamefont
		{Selinger}, \citenamefont {Sun},\ and\ \citenamefont
		{Wei}}]{Chakrabarty2014}%
	\BibitemOpen
	\bibfield  {author} {\bibinfo {author} {\bibfnamefont {A.}~\bibnamefont
			{Chakrabarty}}, \bibinfo {author} {\bibfnamefont {A.}~\bibnamefont {Konya}},
		\bibinfo {author} {\bibfnamefont {F.}~\bibnamefont {Wang}}, \bibinfo {author}
		{\bibfnamefont {J.~V.}\ \bibnamefont {Selinger}}, \bibinfo {author}
		{\bibfnamefont {K.}~\bibnamefont {Sun}}, \ and\ \bibinfo {author}
		{\bibfnamefont {Q.~H.}\ \bibnamefont {Wei}},\ }\href@noop {} {\bibfield
		{journal} {\bibinfo  {journal} {Langmuir}\ }\textbf {\bibinfo {volume}
			{30}},\ \bibinfo {pages} {13844} (\bibinfo {year} {2014})}\BibitemShut
	{NoStop}%
	\bibitem [{\citenamefont {Delong}\ \emph {et~al.}(2015)\citenamefont {Delong},
		\citenamefont {Balboa~Usabiaga},\ and\ \citenamefont {Donev}}]{Delong2015}%
	\BibitemOpen
	\bibfield  {author} {\bibinfo {author} {\bibfnamefont {S.}~\bibnamefont
			{Delong}}, \bibinfo {author} {\bibfnamefont {F.}~\bibnamefont
			{Balboa~Usabiaga}}, \ and\ \bibinfo {author} {\bibfnamefont {A.}~\bibnamefont
			{Donev}},\ }\href@noop {} {\bibfield  {journal} {\bibinfo  {journal} {The
				Journal of Chemical Physics}\ }\textbf {\bibinfo {volume} {143}},\ \bibinfo
		{pages} {144107} (\bibinfo {year} {2015})}\BibitemShut {NoStop}%
	\bibitem [{\citenamefont {Yang}\ and\ \citenamefont
		{Bevan}(2017)}]{Yang:2017ux}%
	\BibitemOpen
	\bibfield  {author} {\bibinfo {author} {\bibfnamefont {Y.}~\bibnamefont
			{Yang}}\ and\ \bibinfo {author} {\bibfnamefont {M.~A.}\ \bibnamefont
			{Bevan}},\ }\href@noop {} {\bibfield  {journal} {\bibinfo  {journal} {The
				Journal of Chemical Physics}\ }\textbf {\bibinfo {volume} {147}},\ \bibinfo
		{pages} {054902} (\bibinfo {year} {2017})}\BibitemShut {NoStop}%
	\bibitem [{\citenamefont {Roosen-Runge}\ \emph {et~al.}(2021)\citenamefont
		{Roosen-Runge}, \citenamefont {Schurtenberger},\ and\ \citenamefont
		{Stradner}}]{Roosen-Runge:2021wo}%
	\BibitemOpen
	\bibfield  {author} {\bibinfo {author} {\bibfnamefont {F.}~\bibnamefont
			{Roosen-Runge}}, \bibinfo {author} {\bibfnamefont {P.}~\bibnamefont
			{Schurtenberger}}, \ and\ \bibinfo {author} {\bibfnamefont {A.}~\bibnamefont
			{Stradner}},\ }\href@noop {} {\bibfield  {journal} {\bibinfo  {journal}
			{Journal of Physics: Condensed Matter}\ }\textbf {\bibinfo {volume} {33}},\
		\bibinfo {pages} {154002} (\bibinfo {year} {2021})}\BibitemShut {NoStop}%
	\bibitem [{\citenamefont {Verweij}\ \emph {et~al.}(2020)\citenamefont
		{Verweij}, \citenamefont {Moerman}, \citenamefont {Ligthart}, \citenamefont
		{Huijnen}, \citenamefont {Groenewold}, \citenamefont {Kegel}, \citenamefont
		{van Blaaderen},\ and\ \citenamefont {Kraft}}]{VerweijPRR2020}%
	\BibitemOpen
	\bibfield  {author} {\bibinfo {author} {\bibfnamefont {R.~W.}\ \bibnamefont
			{Verweij}}, \bibinfo {author} {\bibfnamefont {P.~G.}\ \bibnamefont
			{Moerman}}, \bibinfo {author} {\bibfnamefont {N.~E.~G.}\ \bibnamefont
			{Ligthart}}, \bibinfo {author} {\bibfnamefont {L.~P.~P.}\ \bibnamefont
			{Huijnen}}, \bibinfo {author} {\bibfnamefont {J.}~\bibnamefont {Groenewold}},
		\bibinfo {author} {\bibfnamefont {W.~K.}\ \bibnamefont {Kegel}}, \bibinfo
		{author} {\bibfnamefont {A.}~\bibnamefont {van Blaaderen}}, \ and\ \bibinfo
		{author} {\bibfnamefont {D.~J.}\ \bibnamefont {Kraft}},\ }\href@noop {}
	{\bibfield  {journal} {\bibinfo  {journal} {Phys. Rev. Research}\ }\textbf
		{\bibinfo {volume} {2}},\ \bibinfo {pages} {033136} (\bibinfo {year}
		{2020})}\BibitemShut {NoStop}%
	\bibitem [{\citenamefont {Verweij}\ \emph {et~al.}(2021)\citenamefont
		{Verweij}, \citenamefont {Moerman}, \citenamefont {Huijnen}, \citenamefont
		{Ligthart}, \citenamefont {Chakraborty}, \citenamefont {Groenewold},
		\citenamefont {Kegel}, \citenamefont {van Blaaderen},\ and\ \citenamefont
		{Kraft}}]{Verweij2021}%
	\BibitemOpen
	\bibfield  {author} {\bibinfo {author} {\bibfnamefont {R.~W.}\ \bibnamefont
			{Verweij}}, \bibinfo {author} {\bibfnamefont {P.~G.}\ \bibnamefont
			{Moerman}}, \bibinfo {author} {\bibfnamefont {L.~P.~P.}\ \bibnamefont
			{Huijnen}}, \bibinfo {author} {\bibfnamefont {N.~E.~G.}\ \bibnamefont
			{Ligthart}}, \bibinfo {author} {\bibfnamefont {I.}~\bibnamefont
			{Chakraborty}}, \bibinfo {author} {\bibfnamefont {J.}~\bibnamefont
			{Groenewold}}, \bibinfo {author} {\bibfnamefont {W.~K.}\ \bibnamefont
			{Kegel}}, \bibinfo {author} {\bibfnamefont {A.}~\bibnamefont {van
				Blaaderen}}, \ and\ \bibinfo {author} {\bibfnamefont {D.~J.}\ \bibnamefont
			{Kraft}},\ }\href@noop {} {\bibfield  {journal} {\bibinfo  {journal} {Journal
				of Physics Materials}\ }\textbf {\bibinfo {volume} {4}},\ \bibinfo {pages}
		{035002} (\bibinfo {year} {2021})}\BibitemShut {NoStop}%
	\bibitem [{\citenamefont {Kraft}\ \emph {et~al.}(2013)\citenamefont {Kraft},
		\citenamefont {Wittkowski}, \citenamefont {ten Hagen}, \citenamefont
		{Edmond}, \citenamefont {Pine},\ and\ \citenamefont {Löwen}}]{Kraft2013}%
	\BibitemOpen
	\bibfield  {author} {\bibinfo {author} {\bibfnamefont {D.~J.}\ \bibnamefont
			{Kraft}}, \bibinfo {author} {\bibfnamefont {R.}~\bibnamefont {Wittkowski}},
		\bibinfo {author} {\bibfnamefont {B.}~\bibnamefont {ten Hagen}}, \bibinfo
		{author} {\bibfnamefont {K.~V.}\ \bibnamefont {Edmond}}, \bibinfo {author}
		{\bibfnamefont {D.~J.}\ \bibnamefont {Pine}}, \ and\ \bibinfo {author}
		{\bibfnamefont {H.}~\bibnamefont {Löwen}},\ }\href@noop {} {\bibfield
		{journal} {\bibinfo  {journal} {Physical Review E}\ }\textbf {\bibinfo
			{volume} {88}},\ \bibinfo {pages} {050301} (\bibinfo {year}
		{2013})}\BibitemShut {NoStop}%
	\bibitem [{\citenamefont {Zheng}\ and\ \citenamefont
		{Han}(2010)}]{Zheng:2010wo}%
	\BibitemOpen
	\bibfield  {author} {\bibinfo {author} {\bibfnamefont {Z.}~\bibnamefont
			{Zheng}}\ and\ \bibinfo {author} {\bibfnamefont {Y.}~\bibnamefont {Han}},\
	}\href@noop {} {\bibfield  {journal} {\bibinfo  {journal} {The Journal of
				Chemical Physics}\ }\textbf {\bibinfo {volume} {133}},\ \bibinfo {pages}
		{124509} (\bibinfo {year} {2010})}\BibitemShut {NoStop}%
	\bibitem [{\citenamefont {Hueckel}\ \emph {et~al.}(2021)\citenamefont
		{Hueckel}, \citenamefont {G.M.},\ and\ \citenamefont
		{Sacanna}}]{Hueckel2021}%
	\BibitemOpen
	\bibfield  {author} {\bibinfo {author} {\bibfnamefont {T.}~\bibnamefont
			{Hueckel}}, \bibinfo {author} {\bibfnamefont {H.}~\bibnamefont {G.M.}}, \
		and\ \bibinfo {author} {\bibfnamefont {S.}~\bibnamefont {Sacanna}},\
	}\href@noop {} {\bibfield  {journal} {\bibinfo  {journal} {Nature Review
				Matter}\ }\textbf {\bibinfo {volume} {6}},\ \bibinfo {pages} {1053} (\bibinfo
		{year} {2021})}\BibitemShut {NoStop}%
	\bibitem [{\citenamefont {Glotzer}\ and\ \citenamefont
		{Solomon}(2007)}]{Glotzer:2007uo}%
	\BibitemOpen
	\bibfield  {author} {\bibinfo {author} {\bibfnamefont {S.~C.}\ \bibnamefont
			{Glotzer}}\ and\ \bibinfo {author} {\bibfnamefont {M.~J.}\ \bibnamefont
			{Solomon}},\ }\href@noop {} {\bibfield  {journal} {\bibinfo  {journal}
			{Nature Materials}\ }\textbf {\bibinfo {volume} {6}},\ \bibinfo {pages} {557}
		(\bibinfo {year} {2007})}\BibitemShut {NoStop}%
	\bibitem [{\citenamefont {Fernandez-Rico}\ \emph {et~al.}(2020)\citenamefont
		{Fernandez-Rico}, \citenamefont {Chiappini}, \citenamefont {Yanagishima},
		\citenamefont {de~Sousa}, \citenamefont {Aarts}, \citenamefont {Dijkstra},\
		and\ \citenamefont {Dullens}}]{Fernandez-Rico:2020wu}%
	\BibitemOpen
	\bibfield  {author} {\bibinfo {author} {\bibfnamefont {C.}~\bibnamefont
			{Fernandez-Rico}}, \bibinfo {author} {\bibfnamefont {M.}~\bibnamefont
			{Chiappini}}, \bibinfo {author} {\bibfnamefont {T.}~\bibnamefont
			{Yanagishima}}, \bibinfo {author} {\bibfnamefont {H.}~\bibnamefont
			{de~Sousa}}, \bibinfo {author} {\bibfnamefont {D.~G.~L.}\ \bibnamefont
			{Aarts}}, \bibinfo {author} {\bibfnamefont {M.}~\bibnamefont {Dijkstra}}, \
		and\ \bibinfo {author} {\bibfnamefont {R.~P.~A.}\ \bibnamefont {Dullens}},\
	}\href@noop {} {\bibfield  {journal} {\bibinfo  {journal} {Science}\ }\textbf
		{\bibinfo {volume} {369}},\ \bibinfo {pages} {950} (\bibinfo {year}
		{2020})}\BibitemShut {NoStop}%
	\bibitem [{\citenamefont {Fernandez-Rico}\ and\ \citenamefont
		{Dullens}(2021)}]{Fernandez-Rico:2021uw}%
	\BibitemOpen
	\bibfield  {author} {\bibinfo {author} {\bibfnamefont {C.}~\bibnamefont
			{Fernandez-Rico}}\ and\ \bibinfo {author} {\bibfnamefont {R.~P.~A.}\
			\bibnamefont {Dullens}},\ }\href@noop {} {\bibfield  {journal} {\bibinfo
			{journal} {Proceedings of the National Academy of Sciences}\ }\textbf
		{\bibinfo {volume} {118}},\ \bibinfo {pages} {e2107241118} (\bibinfo {year}
		{2021})}\BibitemShut {NoStop}%
	\bibitem [{\citenamefont {Persat}\ \emph {et~al.}(2014)\citenamefont {Persat},
		\citenamefont {Stone},\ and\ \citenamefont {Gitai}}]{Persat2014}%
	\BibitemOpen
	\bibfield  {author} {\bibinfo {author} {\bibfnamefont {A.}~\bibnamefont
			{Persat}}, \bibinfo {author} {\bibfnamefont {H.~A.}\ \bibnamefont {Stone}}, \
		and\ \bibinfo {author} {\bibfnamefont {Z.}~\bibnamefont {Gitai}},\
	}\href@noop {} {\bibfield  {journal} {\bibinfo  {journal} {Nature
				Communications}\ }\textbf {\bibinfo {volume} {5}},\ \bibinfo {pages} {3824}
		(\bibinfo {year} {2014})}\BibitemShut {NoStop}%
	\bibitem [{\citenamefont {Schuech}\ \emph {et~al.}(2019)\citenamefont
		{Schuech}, \citenamefont {Hoehfurtner}, \citenamefont {Smith},\ and\
		\citenamefont {Humphries}}]{Schuech2019}%
	\BibitemOpen
	\bibfield  {author} {\bibinfo {author} {\bibfnamefont {R.}~\bibnamefont
			{Schuech}}, \bibinfo {author} {\bibfnamefont {T.}~\bibnamefont
			{Hoehfurtner}}, \bibinfo {author} {\bibfnamefont {D.~J.}\ \bibnamefont
			{Smith}}, \ and\ \bibinfo {author} {\bibfnamefont {S.}~\bibnamefont
			{Humphries}},\ }\href@noop {} {\bibfield  {journal} {\bibinfo  {journal}
			{Proceedings of the National Academy of Sciences of the United States of
				America}\ }\textbf {\bibinfo {volume} {116}},\ \bibinfo {pages} {14440}
		(\bibinfo {year} {2019})}\BibitemShut {NoStop}%
	\bibitem [{\citenamefont {Qualmann}\ \emph {et~al.}(2011)\citenamefont
		{Qualmann}, \citenamefont {Koch},\ and\ \citenamefont
		{Kessels}}]{Qualmann2011}%
	\BibitemOpen
	\bibfield  {author} {\bibinfo {author} {\bibfnamefont {B.}~\bibnamefont
			{Qualmann}}, \bibinfo {author} {\bibfnamefont {D.}~\bibnamefont {Koch}}, \
		and\ \bibinfo {author} {\bibfnamefont {M.~M.}\ \bibnamefont {Kessels}},\
	}\href@noop {} {\bibfield  {journal} {\bibinfo  {journal} {EMBO Journal}\
		}\textbf {\bibinfo {volume} {30}},\ \bibinfo {pages} {3501} (\bibinfo {year}
		{2011})}\BibitemShut {NoStop}%
	\bibitem [{\citenamefont {Shen}\ and\ \citenamefont {Shen}(2015)}]{Shen2016}%
	\BibitemOpen
	\bibfield  {author} {\bibinfo {author} {\bibfnamefont {Y.-B.}\ \bibnamefont
			{Shen}}\ and\ \bibinfo {author} {\bibfnamefont {Z.}~\bibnamefont {Shen}},\
	}\href@noop {} {\bibfield  {journal} {\bibinfo  {journal} {Trends in Cell
				Biology}\ }\textbf {\bibinfo {volume} {25}},\ \bibinfo {pages} {59} (\bibinfo
		{year} {2015})}\BibitemShut {NoStop}%
	\bibitem [{\citenamefont {Gray}\ and\ \citenamefont
		{Hancock}(1955)}]{Gray1955}%
	\BibitemOpen
	\bibfield  {author} {\bibinfo {author} {\bibfnamefont {J.}~\bibnamefont
			{Gray}}\ and\ \bibinfo {author} {\bibfnamefont {G.~J.}\ \bibnamefont
			{Hancock}},\ }\href@noop {} {\bibfield  {journal} {\bibinfo  {journal} {The
				Journal of Experimental Biology}\ }\textbf {\bibinfo {volume} {32}},\
		\bibinfo {pages} {802} (\bibinfo {year} {1955})}\BibitemShut {NoStop}%
	\bibitem [{\citenamefont {Rodenborn}\ \emph {et~al.}(2013)\citenamefont
		{Rodenborn}, \citenamefont {Chen}, \citenamefont {Swinney}, \citenamefont
		{Liu},\ and\ \citenamefont {Zhang}}]{Zhang.2013fzd}%
	\BibitemOpen
	\bibfield  {author} {\bibinfo {author} {\bibfnamefont {B.}~\bibnamefont
			{Rodenborn}}, \bibinfo {author} {\bibfnamefont {C.-H.}\ \bibnamefont {Chen}},
		\bibinfo {author} {\bibfnamefont {H.~L.}\ \bibnamefont {Swinney}}, \bibinfo
		{author} {\bibfnamefont {B.}~\bibnamefont {Liu}}, \ and\ \bibinfo {author}
		{\bibfnamefont {H.~P.}\ \bibnamefont {Zhang}},\ }\href {\doibase
		10.1073/pnas.1219831110} {\bibfield  {journal} {\bibinfo  {journal}
			{Proceedings of the National Academy of Sciences of the United States of
				America}\ }\textbf {\bibinfo {volume} {110}},\ \bibinfo {pages} {E338 47}
		(\bibinfo {year} {2013})}\BibitemShut {NoStop}%
	\bibitem [{\citenamefont {Rost}\ \emph {et~al.}(2020)\citenamefont {Rost},
		\citenamefont {Stimatze}, \citenamefont {Egolf},\ and\ \citenamefont
		{Urbach}}]{Rost2020}%
	\BibitemOpen
	\bibfield  {author} {\bibinfo {author} {\bibfnamefont {B.}~\bibnamefont
			{Rost}}, \bibinfo {author} {\bibfnamefont {J.~T.}\ \bibnamefont {Stimatze}},
		\bibinfo {author} {\bibfnamefont {D.~A.}\ \bibnamefont {Egolf}}, \ and\
		\bibinfo {author} {\bibfnamefont {J.~S.}\ \bibnamefont {Urbach}},\
	}\href@noop {} {\bibfield  {journal} {\bibinfo  {journal} {Phys Rev E}\
		}\textbf {\bibinfo {volume} {102}},\ \bibinfo {pages} {023103} (\bibinfo
		{year} {2020})}\BibitemShut {NoStop}%
	\bibitem [{\citenamefont {Lekkerkerker}\ \emph {et~al.}(1992)\citenamefont
		{Lekkerkerker}, \citenamefont {Poon}, \citenamefont {Pusey}, \citenamefont
		{Stroobants},\ and\ \citenamefont {Warren}}]{Lekkerkerker1992}%
	\BibitemOpen
	\bibfield  {author} {\bibinfo {author} {\bibfnamefont {H.~N.~W.}\
			\bibnamefont {Lekkerkerker}}, \bibinfo {author} {\bibfnamefont {W.~C.-K.}\
			\bibnamefont {Poon}}, \bibinfo {author} {\bibfnamefont {P.~N.}\ \bibnamefont
			{Pusey}}, \bibinfo {author} {\bibfnamefont {A.}~\bibnamefont {Stroobants}}, \
		and\ \bibinfo {author} {\bibfnamefont {P.~B.}\ \bibnamefont {Warren}},\
	}\href@noop {} {\bibfield  {journal} {\bibinfo  {journal} {Europhysics
				Letters}\ }\textbf {\bibinfo {volume} {20}},\ \bibinfo {pages} {559}
		(\bibinfo {year} {1992})}\BibitemShut {NoStop}%
\end{thebibliography}

\end{document}